\documentclass[preprint]{aastex}
\shorttitle{Growth of a Vortex Mode}
\shortauthors{Hanawa \& Matsumoto}
\begin{document}

\title{Growth of a Vortex Mode during Gravitational
Collapse Resulting in Type II Supernova}

\author{Tomoyuki Hanawa}
\affil{Department of Astrophysics, School of Science, Nagoya University,
Chikusa-ku, Nagoya 464-8602, Japan}
\email{hanawa@a.phys.nagoya-u.ac.jp}

\and

\author{Tomoaki Matsumoto} 
\affil{Department of Humanity and Environment, Hosei University,
Fujimi, Chiyoda-ku, Tokyo 102-8160, Japan}
\email{matsu@i.hosei.ac.jp}

\begin{abstract}
We investigate stability of a gravitationally collapsing 
iron core against non-spherical perturbation.  The gravitationally
collapsing iron core is approximated by a similarity solution
for dynamically collapsing polytropic gas sphere.  
We find that the similarity solution is unstable against 
non-spherical perturbations. 
The perturbation grows in proportion to 
$ (t \, - \, t _0) ^{-\sigma} $
while the the central density increases in proportion to 
$ (t \, - \, t _0) ^{-2} $.
The growth rate is $ \sigma \, = \, 1/3 \, + \, \ell \,
(\gamma \, - \, 4/3) $, where $ \gamma $ and $ \ell $ 
denote the polytropic index and the parameter $ \ell $ of
the spherical harmonics, $ Y _{\ell} ^m (\theta, \, \varphi) $, 
respectively. The growing perturbation is dominated by 
vortex motion.  Thus it excites global convection during
the collapse and may contribute to material mixing in a type
II supernova.
\end{abstract}

\keywords{gravitation --- 
hydrodynamics --- instabilities --- stars: supernovae}

\section{INTRODUCTION}

Supernova remnants are appreciably aspherical and globally
asymmetric.  The asymmetry of ejecta indicates that
the supernova explosion is highly non-spherical and
contains non-radial flow (see, e.g., the review by 
Goldreich, Lai \& Sahrling 1997 and references therein).  
The non-radial flow in supernova
explosion is suggested also from x-ray and $ \gamma $-ray
observations of SN1987A; one cannot explain early detection of
x-rays and $ \gamma $-rays from SN1987A without invoking
large scale mixing (see, e.g., the review by Bethe 1990 and the
references therein).  The origin of high velocity pulsars may
also be ascribed to asymmetry of supernova explosion
(see, e.g., Burrows \& Hayes 1996 and references therein).

Asymmetry is amplified by the 
Rayleigh-Taylor instability in the early phase  of
supernova explosion (see, e.g., Falk \& Arnett 1973).
According to detailed numerical
simulations, the growth of the Rayleigh-Taylor instability
accounts for matter mixing inferred from observations of
SN1987A if there exists an appropriate seed of the asymmetry 
(Arnett, Fryxell, \& M\"uller 1989; Hachisu et al. 1990; M\"uller,
Fryxell, \& Arnett 1991; Fryxell, Arnett, \&
M\"uller 1991; Nagataki, Shimizu, \& Kato 1998).
Thus it is worth to consider the origin for the seed of 
asymmetry.

Goldreich et al. (1997) discussed possible instabilities
during the core collapse as a source of the seed.
They suggested possibility that dynamically collapsing
core might be unstable against non-radial perturbation
and proposed to study the stability of the similarity
solution of Yahil (1983).  Applying the polytropic equation
of state, $ P \, = \, K \rho ^\gamma $, to pre-supernova core,
he obtained a similarity solution describing collapse of a
spherical iron core.  In this paper we show that his similarity
is indeed unstable against a vortex mode.  The method of 
stability analysis is essentially the same as that of
Hanawa \& Matsumoto (2000) who investigated the bar mode
instability during collapse prior to protostar formation.
Velocity perturbation dominates over density perturbation in 
the vortex mode while both of them have similar amplitude
in the bar mode.  The growth rate of the vortex mode depends
on the wavenumber, $ \ell $, of the spherical harmonics,
$ Y _\ell ^m (\theta, \, \varphi) $.  The vortex mode
grows in proportion to $ \vert t \, - \, t _0 \vert ^{-\sigma} $,
where $ t _0 $ and $ \sigma \, = \, 1/3 \,
+ \, \ell \, (\gamma \, - \, 4/3) $ denote the epoch of
protoneutron star and growth rate, respectively.

In \S 2 we review the similarity solution of Yahil (1983) for
our stability analysis given in \S 3.  We discuss the mechanism
of the vortex mode in \S 4 and implications to type II supernova
in \S 5.  We show the asymptotic
behavior of the similarity solution and perturbation 
in the region very far from the center in Appendix.

\section{SIMILARITY SOLUTION}

For simplicity we consider gas of which equation of state
is expressed by polytrope,
\begin{equation}
P \; = \; K \, \rho ^\gamma \; , \label{polytrope}
\end{equation}
where $ P $ and $ \rho $ denote the pressure and density,
respectively. 
The hydrodynamical equations are then expressed as
\begin{equation}
\frac{\partial \rho}{\partial t} \; + \;
\mbox{\boldmath$\nabla$} \cdot ( \rho \mbox{\boldmath$v$} ) \; = \; 0 \;
\end{equation}
and 
\begin{equation}
\frac{\partial}{\partial t} ( \rho \mbox{\boldmath$v$} )
\; + \; \mbox{\boldmath$\nabla$} P \; + \; \mbox{\boldmath$\nabla$} \cdot
(\rho \mbox{\boldmath$v$} \otimes \mbox{\boldmath$v$}) \;
+ \; \rho \mbox{\boldmath$\nabla$} \Phi \; = \; 0 \; ,
\end{equation}
where $ \mbox{\boldmath$v$} $ and $ \Phi $ denote the
velocity and gravitational potential, respectively.
The gravitational potential is related with the density
distribution by the Poisson equation,
\begin{equation}
\Delta \Phi \; = \; 4 \pi G \rho \; ,
\label{poisson}
\end{equation}
where $ G $ denotes the gravitational constant.

For later convenience, we introduce the zooming coordinates
of Bouquet et al. (1985) to solve equations (\ref{polytrope})
through (\ref{poisson}). 
The zooming coordinates,
$ (\mbox{\boldmath$\xi$}, \, \tau) $, are related with the
ordinary coordinates, $ (\mbox{\boldmath$r$}, \, t) $, by
\begin{equation}
\left(
\begin{array}{c}
\mbox{\boldmath$\xi$} \\
\tau 
\end{array}
\right) \; = \;
\left(
\begin{array}{c}
\displaystyle
\frac{\mbox{\boldmath$r$}}
{c _{\rm 0} \, \vert t \, - \, t _0 \vert } 
\\
- \, \ln \, \vert 1 \, - \, t / t _0 \vert 
\end{array}
\right) \; ,
\end{equation}
where $ c _0 $ denotes a standard sound speed and
is a function of time $ t $.  The symbol, $ t _0 $, denotes an
epoch at the instant of the protoneutron star formation.
The density in the zooming coordinates,
$ \varrho $, is related with that in the ordinary coordinates,
$ \rho $, by
\begin{equation}
\varrho (\mbox{\boldmath$x$}, \, \tau) \; = \;
4 \pi G \, \rho \, (t \, - \, t _0 ) ^2 \; .
\label{density}
\end{equation}
We define the standard sound speed, $ c _0 $, so that it denotes the
sound speed at a given $ t $ when $ \varrho \; = \; 1 $.  Thus 
it is expressed as
\begin{equation}
c _0 \; = \; \sqrt{\gamma K} \, ( 4 \pi G ) ^{(1 \, - \, \gamma) \,
/ \, 2} \, \vert t \, - \, t _0 \vert ^{1 \, - \, \gamma} \; .
\end{equation}
The pressure in the zooming coordinates, $ p $, is related with
the that in the ordinary coordinates, $ P $, by
\begin{equation}
p \; = \; \frac{4 \pi G }{c _0 {}^2} \,
 P \, (t \, - \, t _0) ^2 \; . \label{pressure}
\end{equation}
Substituting equations (\ref{density}) and (\ref{pressure})
into equation (\ref{polytrope}), we obtain the polytrope
relation in the zooming coordinates,
\begin{equation}
p \; = \; \frac{ \varrho ^\gamma }{\gamma} \; .
\label{polytrope2}
\end{equation}
The velocity in the zooming coordinates, 
$ \mbox{\boldmath$u$} $, is defined as
\begin{equation}
\mbox{\boldmath$u$} \; = \;
\frac{\mbox{\boldmath$v$}}{c _0} \; + \;
(2 \, - \, \gamma) \, 
\frac{\mbox{\boldmath$r$}}{c _0 \, \vert t \, - \, t _0 \vert}
\; .  \label{velocity}
\end{equation}
This velocity denotes that with respect to the zooming coordinates,
and includes the apparent motion, the last term 
in equation (\ref{velocity}).
The gravitational potential in the zooming coordinates,
$ \phi $, is related with that in the ordinary coordinates,
$ \Phi $ by
\begin{equation}
\phi \; = \; \frac{\Phi}{c _0 {}^2} \; .
\end{equation}

In the zooming coordinates, the hydrodynamical equations are
expressed as
\begin{equation}
\frac{\partial \varrho}{\partial \tau} \; + \;
\mbox{\boldmath$\nabla$} _\xi \cdot (\varrho \mbox{\boldmath$u$}) \; = \;
(4 \, - \, 3 \gamma ) \, \varrho \; , \label{sim1}
\end{equation}
\begin{equation}
\frac{\partial}{\partial \tau} \, 
(\varrho \mbox{\boldmath$u$}) \, + \,
\mbox{\boldmath$\nabla$} _\xi \cdot
( \varrho \mbox{\boldmath$u$} \otimes \mbox{\boldmath$u$}
) \, + \, \mbox{\boldmath$\nabla$} _\xi p \, + \, \varrho 
\mbox{\boldmath$\nabla$} _\xi \phi
\, = \, (2 \, - \, \gamma)\, (\gamma \, - \, 1) \, 
\varrho \mbox{\boldmath$\xi$} \, + \,
(7 \, - \, 5 \gamma) \, \varrho \mbox{\boldmath$u$} 
\; , \label{sim2}
\end{equation}
and
\begin{equation}
\Delta _\xi \, \phi \; = \; \varrho \;  \label{sim3} 
\end{equation}
for $ t \, < \, t _0 $.
The symbols, $ \mbox{\boldmath$\nabla$} _\xi $ and $ \Delta _\xi $,
denote the gradient and Laplacian in the 
$ \mbox{\boldmath$\xi$} $-space, respectively.

Assuming stationarity in the zooming coordinates 
($ \partial / \partial \tau \; = \; 0 $) and spherical symmetry
($ \partial / \partial \theta \; = \; 
\partial / \partial \varphi \; = \; 0 $), we seek
a similarity solution.  Under these assumptions equations (\ref{sim1}),
(\ref{sim2}), and (\ref{sim3}) reduce to 
\begin{equation}
\frac{\partial u _r}{\partial \xi} \; + \;
\frac{u_r}{\varrho} \, \frac{\partial \varrho}{\partial \xi}
\; = \; (4 \, - \, 3 \gamma) \, - \, \frac{2 u _r}{\xi} \; ,
\label{sim4}
\end{equation}
\begin{equation}
u _r \frac{\partial u _r}{\partial \xi} \; + \;
\frac{1}{\varrho} \, \left( \frac{dp}{d\varrho} \right) \,
\frac{\partial \varrho}{\partial \xi} \; + \;
\frac{\partial \phi}{\partial \xi} \; = \;
(2 \, - \, \gamma) \, (\gamma \, - \, 1) \, \xi \; + \;
(3 \, - \, 2 \gamma) \, u _r \; \; ,
\label{sim5}
\end{equation}
and
\begin{equation}
\frac{\partial \phi}{\partial \xi} \; = \;
\frac{1}{\xi^2} \, \int _0 ^\xi \varrho (\zeta) \, \zeta ^2 \;
d\zeta \; = \; \frac{\varrho u _r}{4 \, - \, 3 \gamma} \; ,
\end{equation}
where $ \xi \, = \, \vert \mbox{\boldmath$\xi$} \vert $.
After some algebra we can rewrite equations (\ref{sim4}) and
(\ref{sim5}) into
\begin{equation}
( \varrho ^{\gamma \, - \, 1} \, - \, u _r {}^2 )
\, \left( \frac{d \varrho}{d \xi} \right) \; = \; \varrho \,
\left\lbrack - \, \frac{\varrho u _r}{4 \, - \, 3 \gamma} \; +
\; (2 \, - \, \gamma) (\gamma \, - \, 1) \, \xi \; + \;
(\gamma \, - \, 1) \, u _r \, + \, \frac{2 u _r {}^2}{\xi} 
\right\rbrack \; , \label{sim6}
\end{equation}
and
\begin{eqnarray}
( \varrho ^{\gamma \, - \, 1} \, - \, u _r {}^2 ) \,
\left( \frac{d u _r}{d\xi} \right) & = &
\frac{ \varrho u _r {}^2 }{4 \, - \, 3 \gamma} \; - \;
(2 \, - \, \gamma) \, (\gamma \, - \, 1) \, \xi u _r \; - \;
(3 \, - \, 2 \gamma) \, u _r {}^2 \; \nonumber \\
& \; & + \; (4 \, - \, 3 \gamma)
\, \varrho ^{\gamma \, - \, 1} \; - \; 
\frac{2 u _r}{\xi} \, \varrho ^{\gamma \, - \, 1} \; .
\label{sim7} 
\end{eqnarray}
Equations (\ref{sim6}) and (\ref{sim7}) are singular at the
sonic point, $ u _r {}^2 \, = \, \varrho ^{\gamma \, - \, 1} $.
We obtain the the similarity solution by integrating equations
(\ref{sim6}) and (\ref{sim7}) with the Runge-Kutta method. 
In the numerical integration we used the auxiliary variable of
Whitworth \& Summers (1985), $ s $, defined by
\begin{equation}
\frac{d \xi}{d s} \; = \; 
\varrho ^{\gamma \, - \, 1} \, - \, u _r {}^2 
\; . \label{WSvariable}
\end{equation}
Using equation (\ref{WSvariable}), we rewrite equations (\ref{sim6})
and (\ref{sim7}) into
\begin{equation}
\frac{d \varrho}{d s} \; = \; \varrho \,
\left\lbrack - \, \frac{\varrho u _r}{4 \, - \, 3 \gamma} \; +
\; (2 \, - \, \gamma) (\gamma \, - \, 1) \, \xi \; + \;
(\gamma \, - \, 1) \, u _r \, + \, \frac{2 u _r {}^2}{\xi} 
\right\rbrack \; , \label{sim8}
\end{equation}
and
\begin{eqnarray}
\frac{d u _r}{d s} & = &
\frac{ \varrho u _r {}^2 }{4 \, - \, 3 \gamma} \; - \;
(2 \, - \, \gamma) \, (\gamma \, - \, 1) \, \xi u _r \; - \;
(3 \, - \, 2 \gamma) \, u _r {}^2 \; \nonumber \\
& \; & + \; (4 \, - \, 3 \gamma)
\, \varrho ^{\gamma \, - \, 1} \; - \; 
\frac{2 u _r}{\xi} \, \varrho ^{\gamma \, - \, 1} \; ,
\label{sim9} 
\end{eqnarray}
respectively.

Similarity solutions exist for $ \gamma \, < \, 4/3 $.
Figure 1 shows the similarity solution for $ \gamma $ =
1.3.  The solid curves denote $ \varrho $
while the dashed curves denote the infall velocity,
$ - v _r \, = \, - \, u _r \, + \, (2 \, - \, \gamma ) \, \xi $.
These solutions are the same as those obtained by Yahil (1983)
and Suto \& Silk (1988).  They have the asymptotic forms of
\begin{equation}
\varrho _0 \; = \; \varrho _{\rm c} \; - \;
\frac{ \varrho _{\rm c} ^{2 \, - \, \gamma} }{6} \,
\left( \varrho _{\rm c} \, - \, \frac{2}{3} \right) 
\, \xi ^2 \; + \; {\cal O} \, (\xi ^4) \; , \label{sol1}
\end{equation}
and
\begin{equation}
\mbox{\boldmath$u$} _0 \; = \; \left\lbrack \left( \frac{4}{3} \, - \, \gamma
\right) \, \xi \; + \; \frac{\varrho _{\rm c} ^{1 \, - \, \gamma}}
{15} \, \left( \varrho _{\rm c} \, - \, \frac{2}{3} \right)
\, \left( \frac{4}{3} \, - \, \gamma \right) \, \xi ^3 \; 
+ \; {\cal O} \, (\xi ^5) \right\rbrack \, 
\mbox{\boldmath$e$} _\xi \; . \label{sol2}
\end{equation}
The value of $ \varrho _{\rm c} $ is 22.04 for $ \gamma $ = 1.3.

\section{VORTEX MODE}

In this section we consider a non-spherical perturbation
around the similarity solution.  
The density perturbation is assumed to be proportional
to the spherical harmonics, $ Y _\ell ^m (\theta, \, \varphi) $.
Then the density and velocity are expressed as 
\begin{eqnarray}
\varrho & = &
\varrho _0 \, + \, \delta \varrho (\xi) \, e ^{\sigma \tau}
\, Y _\ell ^m (\theta, \, \varphi) \; , \label{eigen-density} \\
u _r & = & u _{r0} \, + \, \delta u _r (\xi) \, e ^{\sigma \tau}
\, Y _\ell ^m (\theta, \, \varphi) \, , \label{per1} \\
u _{\theta} & = & \delta u _{\theta} (\xi) \, 
\frac{e ^{\sigma \tau}}{\ell \, + \, 1}
\frac{\partial}{\partial \theta} Y _\ell ^m (\theta, \, \varphi)
\, , \label{per2} \\
u _{\varphi} & = & \delta u _{\theta} (\xi) \, 
\frac{e ^{\sigma \tau}}{(\ell \, + \, 1) \, \sin \, \theta}
\frac{\partial}{\partial \varphi} Y _\ell ^m (\theta, \, \varphi)
\, , \label{per3} \\
\phi & = & \phi _0 \, + \, \delta \phi (\xi) \, e ^{\sigma \tau} \,
Y _\ell ^m (\theta, \, \varphi) \; 
\label{per4} \; ,
\end{eqnarray}
where the symbols with suffix, 0, denote the values in the similarity
solution and the symbols with the symbol, $ \delta $, 
denote the perturbations.
Substituting equations (\ref{per1}) throughout (\ref{per4}) 
into equations (\ref{sim1}), (\ref{sim2}), and
(\ref{sim3}), we obtain the perturbation equations,
\begin{equation}
(\sigma \, + \, 3 \gamma \, - \, 4) \, \delta \varrho \, + \,
\frac{1}{\xi ^2} \, \frac{\partial}{\partial \xi} \,
\lbrack \xi ^2 \, ( \varrho _0 \, \delta u _r \, + \, u _{r0} \delta \varrho) 
\rbrack \, - \, \ell \, \frac{\varrho _0 \delta u _\theta}{\xi}
\, = \, 0 \; , \label{per-1}
\end{equation}
\begin{equation}
(\sigma \, + \, 2 \gamma \, - \, 3) \, \delta u _r
\, + \, \frac{\partial}{\partial \xi} \, 
(u _{r0} \delta u _r ) \, + \,
\frac{\partial}{\partial \xi} \, \left( \frac{ \delta \varrho }
{\rho _0 ^{2 \, - \, \gamma} } \right) \, + \, \delta \Gamma
\, = \, 0 \; , \label{per-2}
\end{equation}
\begin{equation}
(\sigma \, + \, 2 \gamma \, - \, 3) \, \delta u _\theta \, + \,
\frac{u_{r0}}{\xi} \, \frac{\partial}{\partial \xi} \, 
(\xi \delta u _\theta) \, + \,
\frac{\ell \, + \, 1}{\xi} \, \left( \frac{\delta \varrho}
{\varrho _0 {}^{2 \, - \, \gamma}} \, + \, \delta \phi
\right) \, = \, 0 \; , \label{per-3}
\end{equation}
\begin{equation}
\frac{\partial}{\partial \xi} \, \delta \phi \, = \, \delta \Gamma \; ,
\label{per-4}
\end{equation}
and
\begin{equation}
\frac{\partial}{\partial \xi} \, \delta \Gamma \, = \,
- \, \frac{2 \, \delta \Gamma}{\xi} \, + \,
\frac{\ell \, ( \ell \, + \, 1)}{\xi ^2} \, \delta \phi \,
+ \, \delta \varrho \; . \label{per-5} 
\end{equation}
These perturbation equations have singularities at the
origin ($ \xi \, = \, 0 $), the sonic point 
$ \lbrack (u _{r0}) ^2 \, - \, \varrho ^{\gamma \, - \, 1} 
\rbrack $, and the infinity ($ \xi \, = \, + \infty $).
These perturbation equations
are the same as those of Hanawa \& Matsumoto (2000).

To obtain the boundary condition at the origin we use
the Taylor expansion of the perturbation.
In the following we use the notation,
\begin{eqnarray}
\frac{\delta \varrho}{\varrho _0 {}^{2 \, - \, \gamma}}
& = & \varepsilon _0 \, \xi ^\ell \; + \; 
\varepsilon _2 \, \xi ^{\ell \, + \, 2} \; + \;
{\cal O} \, (\xi ^{\ell \, + \, 4} ) \; , \label{per-density} \\
\delta u _r & = & \alpha _0 \, \xi ^{\ell \, - \, 1} \; + \;
\alpha _2 \, \xi ^{\ell \, + \, 1} \; + \;
{\cal O} \, (\xi ^{\ell \, + \, 3} ) \; , \\
\delta u _\theta & = & \beta _0 \, \xi ^{\ell \, - \, 1} \; + \;
\beta _2 \, \xi ^{\ell \, + \, 1} \; + \;
{\cal O} \, (\xi ^{\ell \, + \, 3} )  \; ,
\end{eqnarray}
and
\begin{equation}
\delta \phi \; = \; \lambda _0 \, \xi ^\ell \; + \; \lambda _2 \,
\xi ^{\ell \, + \, 2} \; + \; {\cal O} \, (\xi ^{\ell \, + \, 4}) \; .
\label{dpotential}
\end{equation}
Substituting equation (\ref{dpotential}) into equation 
(\ref{per-4}) we obtain
\begin{equation}
\delta \Gamma \; = \; \ell \, \lambda _0 \, \xi ^{\ell \, - \, 1} \; + \;
(\ell \, + \, 2) \, \lambda _2 \, \xi ^{\ell \, + \, 1} \; + \;
{\cal O} \, (\xi ^{\ell \, + \, 3}) \; . \label{dgravity} 
\end{equation}
Note that not only the leading terms but the second lowest
terms are taken into account in equations (\ref{per-density})
through (\ref{dgravity}).  

Substituting equations 
(\ref{per-density}) through (\ref{dgravity}) into equations
(\ref{per-1}) through (\ref{per-5}) we derive conditions
for $ \alpha _0 $,  $ \alpha _2 $, $ \beta _0 $,
$ \beta _2 $, $ \varepsilon _0 $, 
$ \varepsilon _2 $, $ \lambda _0 $, and $ \lambda _2 $.
From equation (\ref{per-1}) we obtain
\begin{equation}
(\ell \, + \, 1) \, \alpha _0 \; - \; \ell \, \beta _0 \; = \; 0 \; .
\label{per-1-1}
\end{equation}
Equation (\ref{per-1-1}) ensures that the terms proportional
to $ \xi ^{\ell \, - \, 2} $ vanish in equation (\ref{per-1}). 
The terms proportional to $ \xi ^{\ell \, - \, 2} $ vanish
in equation (\ref{per-5}) at any condition.
From equations (\ref{per-2}) and (\ref{per-3}) we obtain
\begin{equation}
\left\lbrack \sigma \, + \, 2 \gamma \, - \, 3 \, + \,
\ell \, \left( \frac{4}{3} \, - \, \gamma \right) \right\rbrack
\, \alpha _0 \; + \; \ell \, (\varepsilon _0 \, + \, \lambda _0)
\; = \; 0 \; , \label{per-2-1}
\end{equation}
and
\begin{equation}
\left\lbrack \sigma \, + \, 2 \gamma \, - \, 3 \, + \,
\ell \, \left( \frac{4}{3} \, - \, \gamma \right) \right\rbrack
\, \beta _0 \; + \; (\ell \, + \, 1) \, (\varepsilon _0 \, + \, \lambda _0)
\; = \; 0 \; ,
\end{equation}
respectively.  These equations ensure that the terms proportional
to $ \xi ^{\ell \, - \, 1} $ vanish in equations (\ref{per2})
and (\ref{per3}).  Similarly we obtain 
\begin{eqnarray}
& \; & \varrho _{\rm c} ^{2 \, - \, \gamma} \,
\left\lbrack \sigma \, + 
\ell \, \left( \frac{4}{3} \, - \, \gamma \right)
\right\rbrack \, \varepsilon _0 
\; + \; 
\varrho _{\rm c} \, \lbrack (\ell \, + \, 3) \, \alpha _2 \,
- \, \ell \, \beta _2 \rbrack \nonumber \\
& \; & - \; \frac{ \varrho _{\rm c} {}^{2 \, - \, \gamma}}{6} \,
\left( \varrho _{\rm c} \, - \, \frac{2}{3} \right) \,
\left\lbrack (\ell \, + \, 3) \, \alpha _0 \, - \, \ell \,
\beta _0 \right\rbrack \; = \; 0 \; ,
\end{eqnarray}
and
\begin{equation}
\lbrack (\ell \, + \, 2) \, (\ell \, + \, 3) \, - \,
\ell \, (\ell \, + \, 1) \rbrack \, \lambda _2 \; = \;
\varrho _{\rm c} {} ^{2 \, - \, \gamma} \, \varepsilon _0 \; , 
\end{equation}
from the condition that the terms proportional to 
$ \xi ^\ell $ vanish and
\begin{eqnarray}
\left\lbrack \sigma \, + \, 2 \gamma \, - \, 3 \, + \,
(\ell \, + \, 2) \, \left( \frac{4}{3} \, - \, \gamma \right)
\right\rbrack \, \alpha _2 & + & (\ell \, + \, 2) \,
\frac{\varrho _{\rm c} {}^{1 \, - \, \gamma}}{15} \,
\left( \varrho _{\rm c} \, - \, \frac{2}{3} \right) \,
\left( \frac{4}{3} \, - \, \gamma \right) \, \alpha _0 
\nonumber \\ 
& + & (\ell \, + \, 2) \, (\varepsilon _2 \, + \, \lambda _2) 
\; = \; 0 \; , \label{high1}
\end{eqnarray}
and
\begin{eqnarray}
\left\lbrack \sigma \, + \, 2 \gamma \, - \, 3 \, + \,
(\ell \, + \, 2) \, \left( \frac{4}{3} \, - \, \gamma \right)
\right\rbrack \, \beta _2 & + & \ell \,
\frac{\varrho _{\rm c} {}^{1 \, - \, \gamma}}{15} \,
\left( \varrho _{\rm c} \, - \, \frac{2}{3} \right) \,
\left( \frac{4}{3} \, - \, \gamma \right) \, \beta _0 
\nonumber \\
& + & (\ell \, + \, 1) \, (\varepsilon _2 \, + \, \lambda _2) 
\; = \; 0 \label{high2}
\end{eqnarray}
from the condition that the terms proportional to 
$ \xi ^{\ell \, + \, 1} $ vanish.

From equations  (\ref{per-1-1}), (\ref{high1}) and (\ref{high2}) 
we obtain
\begin{equation}
\left\lbrack \sigma \, + \, 2 \gamma \, - \, 3 \, + \,
(\ell \, + \, 2) \, \left( \frac{4}{3} \, - \, \gamma \right) 
\right\rbrack \,
\lbrack (\ell \, + \, 1) \, \alpha _2 \; - \; 
(\ell \, + \, 2) \, \beta _2 \rbrack \; = \; 0 \; .
\end{equation}
This condition is equivalent to the condition that
either of 
\begin{equation}
(\ell \, + \, 1) \, \alpha _2 \; - \; (\ell \, + \, 2) \,
\beta _2 \; = \; 0 \label{high3}
\end{equation}
and
\begin{equation}
\sigma \; = \; \frac{1}{3} \; + \; \ell \,
\left( \gamma \, - \, \frac{4}{3} \right) \label{high4}
\end{equation}
is fulfilled.   The bar mode found by Hanawa \& Matsumoto (2000)
fulfills equation (\ref{high3}).  We seek the other mode that
fulfills equation (\ref{high4}).  In the following we call the
latter the vortex mode.

Since the perturbation is linear, the solution can be 
arbitrarily scaled.  To normalize the solution we
take $ \alpha _0 \; = \; 1 $ in this paper.
Then we obtain
\begin{equation}
\beta _0 \; = \; \frac{\ell \, + \, 1}{\ell} \; ,
\end{equation}
\begin{equation}
\lambda _0 \; = \; - \, \varepsilon _0 \; + \;
\frac{2}{\ell} \, \left( \frac{4}{3} \, - \, \gamma \right) \; ,
\label{implication}
\end{equation}
\begin{equation}
\varepsilon _0 \; = \; - 3 \,
\varrho _{\rm c} ^{\gamma \, - \, 1} \, 
\lbrack (\ell \, + \, 3) \, \alpha _2
\, - \, \ell \, \beta _2 \rbrack \; + \;
\left( \varrho _{\rm c} \, - \, \frac{2}{3} \right) \; ,
\end{equation}
\begin{equation}
\lambda _2 \; = \; \frac{\varrho _{\rm c} ^{2 \, - \, \gamma}}
{2 \, (2 \ell \, + \, 3)} \, \varepsilon _0 \; ,
\end{equation}
and
\begin{equation}
\varepsilon _2 \; = \; - \,
\frac{\varrho _{\rm c} ^{1 \, - \, \gamma}}{15}
\, \left( \varrho _{\rm c} \, - \, \frac{2}{3} \right)
\, \left( \frac{4}{3} \, \, - \, \gamma \right) \; - \;
\lambda _2 \; .
\end{equation}

We obtained an eigenfunction of this vortex mode numerically
by the following procedures.  First we integrated
equations (\ref{sim8}) and (\ref{sim9}) to obtain the
similarity solution for a given $ \gamma $.  Second we 
obtained three linearly independent solutions for the
perturbation around the origin by using the Taylor
series expansion, equations (\ref{per-density}) through
(\ref{dgravity}).  Third we integrated the three 
linearly independent solutions from the origin toward the
sonic point numerically with the Runge-Kutta method.
Forth we obtained two linearly independent solutions satisfying
the boundary conditions both at the origin and sonic point
by taking linear combinations of the numerically integrated
solutions.  Finally we integrated the two linearly independent
solutions from the sonic point toward the infinity and
obtained an eigenfunction satisfying all the boundary conditions
as a linear combination of them.  The eigenfunction should
have an infinitesimal small amplitude at the infinity as shown
in Appendix.

Figures 2, 3, and 4 denote the numerically obtained eigenfunctions 
of $ \ell \, = \, 1 $, 2, and 3, respectively.
The polytropic index is set to be $ \gamma $ = 1.3 in the figures.
The eigenfunctions are normalized so that the radial velocity
perturbation is $ \delta u _r \, = \, \xi ^{\ell \, - \, 1} \,
+ \, {\cal O} \, (\xi ^{\ell \, + \, 1}) $ near the origin.
The non-radial velocity perturbation, $ \delta u _\theta $,
changes its sign around $ \xi \, \approx \, 1 $.
The density perturbation is small.  It should be proportional
to $ (\gamma \, - \, 4/3) $ [see eq. (\ref{implication})].

The vortex mode of $ \ell $ = 1 is different from the ghost mode
of $ \ell $ = 1 (Hanawa \& Matsumoto 1999, 2000).  The vortex mode
denotes circulation streaming back through the core surface while
the ghost mode denotes misfit of the gravity center to the
coordinate center (Hanawa \& Matsumoto 1999).  The ghost mode
has no vortex ($ \mbox{\boldmath$\nabla$} _\xi 
\mbox{\boldmath$\times$} \delta \mbox{\boldmath$u$} \, = \, 0 $). 
The vortex mode of $ \ell $ = 1 has
the growth rate of $ \sigma $ = $ \gamma \, - \, 1 $ while
the ghost mode has that of $ \sigma $ = $ 2 \, - \, \gamma $.

Figure 5 shows the density and velocity perturbation in
the cross section.  The contours denote the iso-density curves
of $ \varrho $ = 10, 3, 1, 0.3, and 0.1 
while the central density is $ \varrho _{\rm c} $ = 22.04.
The arrows denote the velocity vectors. 
We obtained the 
density and velocity by adding the eigenmode of $ (\ell, \, m) $ = 
(2,~0) on the similarity solution for $ \gamma $ = 1.3. 
The infall is a little faster in the $ x $-direction than
in the $ z $-direction.  
Figure 6 is the same as Figure 5
but only the velocity perturbation is shown by the arrows in
Figure 6.  It shows that this eigenmode is 
an vortex flow in the meridional plane. 

\section{GROWTH MECHANISM}

In this section we discuss the mechanism for the growth of the
vortex mode.  We consider the conservation
of vorticity to elucidate the growth mechanism.  
Taking rotation of the equation of motion,
we obtain,
\begin{equation}
\frac{\partial \mbox{\boldmath$\Omega$}}{\partial t}
\; = \; \mbox{\boldmath$\nabla$} \, \mbox{\boldmath$\times$} \, 
(\mbox{\boldmath$v$} \mbox{\boldmath$\times$}
\mbox{\boldmath$\Omega$}) \; , \label{vortex-ordinary} 
\end{equation}
where
\begin{equation}
\mbox{\boldmath$\Omega$} \; = \; \mbox{\boldmath$\nabla$}
\mbox{\boldmath$\times$} \mbox{\boldmath$v$}
\; ,
\end{equation}
since the rotation of the pressure force and that of gravity
vanish.  Similarly the conservation of the vorticity is
expressed as
\begin{equation}
\frac{\partial \mbox{\boldmath$\omega$}}{\partial \tau}
\; = \; \mbox{\boldmath$\nabla$} _\xi \, \mbox{\boldmath$\times$}
 \, (\mbox{\boldmath$u$} \mbox{\boldmath$\times$}
\mbox{\boldmath$\omega$}) \; + \;
( 3 \, - \, 2 \gamma ) \,
\mbox{\boldmath$\omega$} \; , \label{vortex-zooming} 
\end{equation}
where
\begin{equation}
\mbox{\boldmath$\omega$} \; = \; 
\mbox{\boldmath$\nabla$} _\xi \,  \mbox{\boldmath$\times$} 
\, \mbox{\boldmath$u$} \; .
\end{equation}
Since $ \mbox{\boldmath$\nabla$} _\xi \mbox{\boldmath$\times$} 
\mbox{\boldmath$u$} _0 \; = \; 0 $, we obtain
\begin{equation}
\frac{\partial}{\partial \tau} \, \delta \mbox{\boldmath$\omega$}
\; = \; \mbox{\boldmath$\nabla$} _\xi \mbox{\boldmath$\times$} \, 
(\mbox{\boldmath$u$} _0 \, \mbox{\boldmath$\times$}
\, \delta \mbox{\boldmath$\omega$} ) \; + \;
(3 \, - \, 2 \gamma) \, \delta \mbox{\boldmath$\omega$} \; ,
\label{per-vortex}
\end{equation}
for the perturbation of the vortex, $ \delta \mbox{\boldmath$\omega$} $.
Substituting the equation (\ref{sol2}) into equation (\ref{per-vortex}) we
obtain
\begin{equation}
\frac{\partial}{\partial \tau} \, \delta \mbox{\boldmath$\omega$}
\; = \; \left\lbrack \left( \gamma \, - \, \frac{4}{3} \right) \,
\frac{\partial}{\partial \ln \, \xi} \; + \; \frac{1}{3} 
\right\rbrack \, \delta \mbox{\boldmath$\omega$} \; ,
\label{vortex}
\end{equation}
near $ \xi \, = \, 0 $.
We can derive equation (\ref{high4}) from equation 
(\ref{vortex}) since $ \delta \mbox{\boldmath$\omega$} \,
\propto \, \xi ^\ell $ near $ \xi \, = \, 0 $ in the vortex mode.  
We can derive also the growth rate of the spin-mode, 
$ \sigma \, = \, 1/3 $,
from equation (\ref{vortex}) since 
$ \delta \mbox{\boldmath$\omega$} \, = \, \xi ^0 $
near $ \xi \, = \, 0 $ in the spin-up mode.
The growth of the vortex mode as well as that of the
spin-up mode is subject to the conservation of the vorticity.

\section{IMPLICATION TO TYPE II SUPERNOVA} 

As shown in the previous sections the vortex mode grows in
proportion to $ \vert t \, - \, t _0 \vert ^{-\sigma} $.
In other words it grows in proportion to 
$ \rho _{\rm c} ^{\sigma / 2} $ since $ \rho _{\rm c}
\, \propto \, (t \, - \, t _0 ) ^{-2} $.  This growth
rate is for the growth of the relative amplitude, i.e.,
that for $ \delta u \, = \, \delta v / c _0 $.  The growth
of the anisotropic velocity is proportional to
\begin{equation}
\delta v \; \propto \; c _0 \, \rho _{\rm c} ^{\sigma/2}
\; \propto \; \rho _{\rm c} ^{\sigma ^\prime} \; ,
\end{equation}
where
\begin{equation}
\sigma ^\prime \; = \; \frac{1}{3} \; + \; \frac{\ell \, + \, 1}{2} \,
\left( \gamma \, - \, \frac{4}{3} \right) \; .
\end{equation}
Suppose that
an iron core during implosion can be well approximated by
a polytrope of $ \gamma \, = \, 1.3 $.  Then the velocity
perturbation of $ \ell \, = \, 1 $ grows by a factor of 30
while the density increases from $ \rho _{\rm c} \, = \,
10 ^9 $ g~cm$^{-3}$ to $ 10 ^{14} $ g~cm$^{-3}$.
Similarly that of $ \ell \, = \, 2 $ grows by a factor
of 25 during the same period.  

If the $ \ell \, = \, 1 $ mode is amplified, the central core, i.e.,
the protoneutron star has a bulk velocity relative to
the envelope.  This might explain a run-away pulsar from its
natal nebula.
If the $ \ell \, \ge \, 2 $ mode is amplified,   
the  anisotropic 
velocity will cause global mixing during the supernova
explosion ($ t \, > \, t _0 $).  This may explain heavy
element exposure earlier than expected from a spherical
symmetric model.

It should be also noted that the vortex mode of 
$ (\ell, \, m) $ = $ (2, \, 0) $ has velocity field similar
to that of the Eddington-Sweet meridional circulation. 
If a pre-supernova star rotates slowly, the meridional
circulation can be a seed of the vortex mode amplified
during the implosion phase.  Convection during Si-burning
may also be the seed for the vortex mode.

\acknowledgments

We thank Shigehiro Nagataki, Katsuhiko Sato, and Shoichi Yamada
for helpful discussion.
This research is financially supported
in part by the Grant-in-Aid for
Scientific Research on Priority Areas of 
the Ministry of Education, Science, Sports and Culture of Japan
(No. 10147105, 11134209).

\appendix

\section{Asymptotic Behavior around the Infinity}

In this appendix we derive asymptotic forms of
perturbations around the similarity solution for a collapsing 
gas sphere.
In the region of $ \xi \, \gg \, 1 $ the similarity
solution has the asymptotic form of
\begin{equation}
\varrho \; \propto \; 
\xi ^{- \, 2 \, / \, (2 \, - \, \gamma) } \; , \label{sol3} 
\end{equation}
and
\begin{equation}
\lbrack u _r \, - \, (2 \, - \, \gamma) \, \xi \rbrack \;
\propto \; \xi ^{(1 \, - \, \gamma) \, / \, (2 \, - \, \gamma)} 
\; . \label{sol4}
\end{equation}
See Yahil (1983) and Suto \& Silk (1988) for the derivation.

As a boundary condition we assume that the relative density
perturbation, $ \delta \varrho / \varrho _0 $, 
is vanishingly small at infinity, 
$ \xi \, = \, \infty $.  After some algebra we obtain the asymptotic 
relations,
\begin{equation}
\frac{\delta \varrho}{\varrho _0} \; \propto \;
\xi ^{- \sigma /(2 \, - \, \gamma)} 
\end{equation}
\begin{equation}
\delta u _r \; \propto \;
\xi ^{- (\sigma \, + \, \gamma \, - \, 1)/(2 \, - \, \gamma)} \; ,
\end{equation}
\begin{equation}
\delta u _\theta \; \propto \;
\xi ^{- (\sigma \, + \, \gamma \, - \, 1)/(2 \, - \, \gamma)} \; ,
\end{equation}
\begin{equation}
\phi \, \propto \, \xi ^{-(\sigma \, - \, 2 \gamma \, + \, 2)
/(2 \, - \, \gamma)} \; ,
\end{equation}
and
\begin{equation}
\phi \, = \, \left\lbrack \frac{(\sigma \, - \, 2 \gamma \, + \, 2)
(\sigma \, - \, 3 \gamma \, + \, 4)}{(2 \, - \, \gamma)^2}
\, - \, \ell \, (\ell \, + \, 1) \right\rbrack ^{-1}
\, r ^2 \, \delta \varrho \; .
\end{equation}
When we derive the above relations, we use equations
(\ref{sol3}) and (\ref{sol4}).  See also Hanawa \& Matsumoto  
(2000) for the derivation.

\newpage

\begin{figure}
\epsscale{0.45}
\plotone{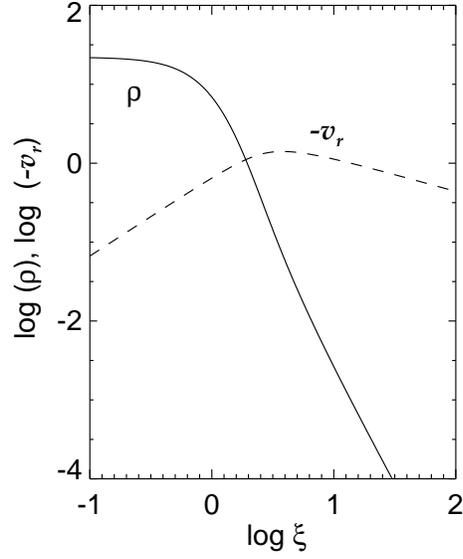} 
\caption{The similarity solution is shown for a collapsing
polytropic gas sphere of $ \gamma $ = 1.30.  The solid curves
denote the density, $ \varrho \, (\xi) $, and the infall
velocity, $ - v _r ( \xi ) \, \equiv \, (2 \, - \, \gamma) \, \xi 
\, - \, u _r \, ( \xi ) $.   The latter is normalized by the
sound speed at the center.}
\end{figure}

\begin{figure}
\epsscale{0.45}
\plotone{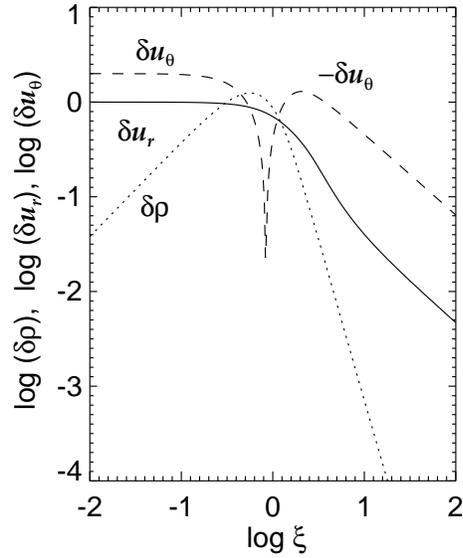} 
\caption{The eigenfunction of $ \ell \, = \, 1 $ mode
is shown as a function of $ \xi $.  The solid curve denotes
the radial velocity perturbation, $ \delta u _r $.  The
dashed and dotted curves denote $ \delta u _\theta $ and
$ \delta \rho $, respectively.}
\end{figure}

\begin{figure}
\epsscale{0.45}
\plotone{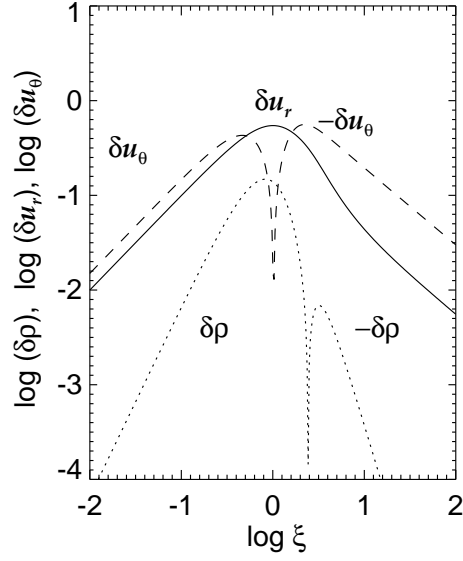} 
\caption{The same as Figure 1 but for $ \ell \, = \, 2 $ mode.}
\end{figure}

\begin{figure}
\epsscale{0.45}
\plotone{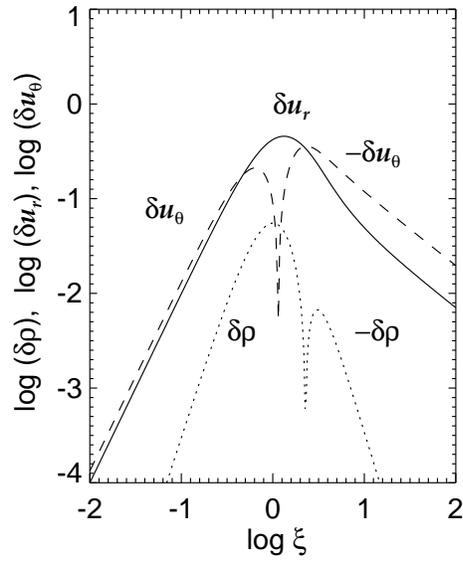} 
\caption{The same as Figure 2 but for $ \ell \, = \, 3 $ mode.}
\end{figure}

\begin{figure}
\epsscale{0.45}
\plotone{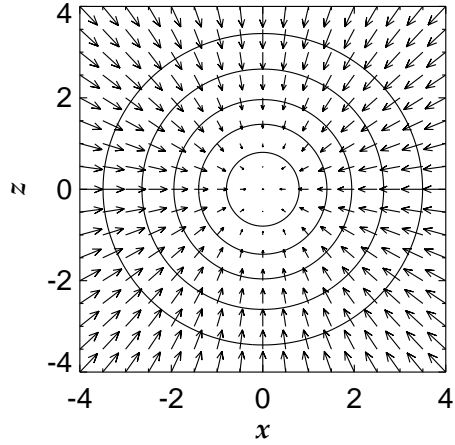} 
\caption{This cross section shows a dynamically collapsing
iron core suffering the vortex mode of $ \ell $ = 2.
The contours denote the isodensity curves of
$ \varrho $ = 10.0, 3.0, 1.0, 0.3, and 0.1.
The arrows denote the velocity, $ \mbox{\boldmath$v$} $,
in the $ x \, - \, z $ plane.}
\end{figure}

\begin{figure}
\epsscale{0.45}
\plotone{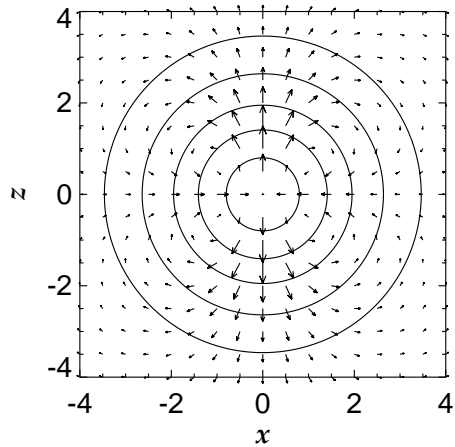} 
\caption{The same as Figure 4 but only the velocity perturbation
is shown by the arrows.}
\end{figure}

\end{document}